\documentclass[conference]{IEEEtran}
\IEEEoverridecommandlockouts
\usepackage{cite}
\usepackage{amsmath,amssymb,amsfonts}
\usepackage{algorithmic}
\usepackage{graphicx}
\usepackage{textcomp}
\usepackage{xcolor}
\usepackage{hyperref}
\usepackage[capitalise]{cleveref}
\usepackage{multirow}
\usepackage{siunitx}
\crefname{algocf}{alg.}{algs.}
\crefname{algocf}{Algorithm}{Algorithms}
\usepackage{todonotes}
\usepackage{siunitx}
\usepackage[ruled,linesnumbered]{algorithm2e}
\newcommand{\NOR}{\texttt{NOR}}

\newcommand{\VDD}{V_{DD}}
\newcommand{\GND}{\textit{GND}}

\def\BibTeX{{\rm B\kern-.05em{\sc i\kern-.025em b}\kern-.08em
    T\kern-.1667em\lower.7ex\hbox{E}\kern-.125emX}}
\begin{document}

\title{Signal Prediction for Digital Circuits by Sigmoidal Approximations using Neural Networks
\\\thanks{This work was conducted in the context of the Austrian Science Fund (FWF) project DMAC (10.55776/P32431).}
}

\author{\IEEEauthorblockN{Josef Salzmann}
\IEEEauthorblockA{\textit{Institute of Computer Engineering, TU Wien} \\
\textit{Vienna, Austria}\\
josef.salzmann@tuwien.ac.at}
\and
\IEEEauthorblockN{Ulrich Schmid}
\IEEEauthorblockA{\textit{Institute of Computer Engineering, TU Wien} \\
\textit{Vienna, Austria}\\
s@ecs.tuwien.ac.at}
}

% TODO: For de-anonymization add github link, authors and FWF reference
% \author{\IEEEauthorblockN{Anonymous}}

\maketitle

\begin{abstract}

Investigating the temporal behavior of digital circuits is a crucial step in system design, usually done via analog or digital simulation. Analog simulators like SPICE iteratively solve the differential equations characterizing the circuits’ components numerically. Although unrivaled in accuracy, this is only feasible for small designs, due to the high computational effort even for short signal traces. Digital simulators use digital abstractions for predicting the timing behavior of a circuit. Besides static timing analysis, which performs corner-case analysis of critical path delays only, dynamic timing analysis provides per-transition timing information in signal traces. In this paper, we advocate a novel approach, which generalizes digital traces to traces consisting of sigmoids, each parameterized by threshold crossing time and slope. What is needed to compute the output trace of a gate is a transfer function, which determines the parameters of the output sigmoids given the parameters of the input sigmoids. Harnessing the power of artificial neural networks (ANN), we implement such transfer functions via ANNs. Using inverters and \NOR\ as the elementary gates in a prototype implementation of a specifically tailored simulator, we demonstrate that our approach operates substantially faster than an analog simulator, while offering better accuracy than a digital simulator.
\end{abstract}

% \begin{IEEEkeywords}
% keywords
% \end{IEEEkeywords}

\section{Introduction and Overview}
\label{sec:intro}
The golden standard for accurate dynamic timing analysis of digital circuits are analog simulators like SPICE \cite{nagel1973spice}. In contrast to static timing analysis, which conducts corner-case analysis of critical path delays in a circuit only, dynamic timing analysis provides per-transition timing information in digital signal traces. By iteratively solving the differential equations that govern the electrical behavior of the transistors of a circuit numerically, highly accurate traces of the output signal waveforms are computed. The drawback of this kind of analysis is the high computational effort and, hence, the low scalability in terms of the circuit size, which makes it feasible for small designs and short traces only. By contrast, digital timing simulators completely abstract away analog waveforms, by discretizing those via zero-time (Heaviside) transitions generated at threshold crossing times. Instead of solving differential equations at simulation time, digital timing analysis tools like ModelSim use pre-computed delay models to parameterize pure or inertial delay channels \cite{Pure_Inertial_Model}, which ultimately determine the digital output signal traces. Prominent examples of such pre-computed delay models are the \emph{Current Source Models} (CSM), the \emph{Effective CSM} (ECSM) \cite{ECSM} and the \emph{Composite CSM} (CCSM) \cite{CCS}. All these models rely on analog simulations for generating tables of voltage (ECSM) or current (CCSM) traces for different input signal slopes and output capacitances.

Given the inability of pure and inertial delays to model pulse degradation, which severely limits the achievable accuracy, several alternative channel models have been proposed in the past. All these models are single history models, where the input-to-output delay $\delta(T)$ of a given input transition depends on the time difference $T$ of the input to the previous output transition. By choosing a suitable function for $\delta(.)$, pulse degradation (up to complete canceling) can be modeled. For example, the authors of the \emph{Delay Degradation Model} (DDM) \cite{Delay_Degradation_Model, DDM_Book} proposed a suitably parametrized exponential function for $\delta(.)$. However, in \cite{FNS16:ToC}, it has been shown that all the existing delay models, including the DDM, do not match reality in some situations. This problem has been avoided by the \emph{Involution Delay Model} (IDM) proposed in \cite{fugger2019faithful} later on, which is based on delay functions that are self-inverse, more specifically, negative involutions satisfying $-\delta(-\delta(T))=T$. And indeed, it has been shown in \cite{NFNS15:GLSVLSI,ohlinger2021involution} that this model surpasses all popular alternative models also in terms of accuracy.

\subsection{Our contributions}
\label{sec:contrib}

Inherently, no digital model, including the IDM, can express and utilize slope information: a Heaviside transition does not reveal whether the corresponding analog waveform has crossed the threshold voltage steeply or not. This might severely change the delay to the next output transition, however.
%Consequently, digital timing simulators cannot even utilize the slope-related information provided in models CSM and ECSM.
In this paper, we propose an approach that mitigates this shortcoming by replacing traces consisting of sequences of Heaviside transitions by traces consisting of sequences of sigmoids, called sigmoidal approximations. Given that a single sigmoid (see \cref{eq:single_sigmoid_equation} for a concrete example) is parametrized with a threshold crossing time, analogous to a Heaviside transition, but also by a slope parameter, this is a very natural generalization. Since every analog waveform can be represented by a sum of appropriately parametrized sigmoid functions \cite{Cyb89}, one may safely expect that timing analysis based on sigmoidal approximations significantly outperforms digital timing analysis in terms of accuracy.

What is needed for dynamic timing analysis here is a way to compute the parameters of the sigmoids generated at some output based on the parameters of the sigmoids on the relevant inputs. For example, analogous to single-history digital channels, one could strive to compute the transition time and the slope of the next output sigmoid of an inverter as functions of the parameters of the current input sigmoid and the previous output sigmoid. Since the transfer functions needed here are considerably more complex than the simple delay functions $\delta(T)$ used in digital models, however, we are still working on the problem of finding reasonably simple but accurate analytic solutions. In order to demonstrate the principal suitability of our sigmoidal approach, which
is the primary purpose of this paper, we resort to using artificial neural networks (ANN) for implementing gate transfer functions as an alternative solution.

In more detail, we developed a prototype of a specifically tailored simulator for circuits composed of inverters and
\NOR\ gates. Their transfer functions were implemented by relatively simple ANNs, which have been trained by sigmoidal
approximations of SPICE-generated analog sample traces of these gates. A number of experiments
using our simulator confirm that it operates substantially faster than an analog simulator, while offering considerably better accuracy than any digital simulator.

\subsection{Related work}
A simple way of constructing involution delay functions, already introduced in \cite{fugger2019faithful}, is a thresholded hybrid model \cite{FFNS23:HSCC}. In the simplest case of a single-input single-output gate like a buffer or inverter, it assembles an internal analog output signal by pasting together pieces of rising and falling switching waveforms in a continuous way. The transitions on the digital input signal dictate the switching instants, and a threshold voltage comparator is used for generating the digital output signal from the analog one. Thresholded hybrid models can also be generalized to multi-input gates \cite{FMOS22:DATE,FSS23:ICCAD}, which allows to also cover \emph{multi-input switching} (MIS) effects. A simple variant of such a model, based on exponential switching waveforms, is also used in the parallelized timing simulator introduced in \cite{schneider2014data}. A fairly old approach for digital dynamic timing simulation, which shares some similarities with simple hybrid models, is IRSIM \cite{SH89:DAC}. Like \cite{FMOS22:DATE,schneider2014data}, the models used there consider transistors as zero-time switches, describe the resulting system as an RC network, and discretize analog switching waveforms using a comparator. Rather than employing continuous mode switching, i.e., pasting together pieces of switching waveforms in a continuous way, however, IRSIM utilizes non-continuous mode switching.

Since the traces generated by all the above models are purely digital, however, they can neither express nor utilize
slope information. Indeed, this is true for all the approaches for digital logic simulation surveyed
in \cite{gunes2005survey, roessler2019survey}.
So, somewhat surprisingly, albeit sigmoids are very popular in other fields of science like cell biology, to the best of our knowledge, they have not been used for modeling signals in digital integrated circuits so far.

It is important to note that our goal is very different from closely approximating small parts of real switching waveforms by suitably parametrized functions. For example, in \cite{MFNS19:ASYNC}, a cubic polynomial was used for well-approximating a single transition of a real switching waveform. By contrast, we are interested in reasonably approximating an arbitrary switching waveform by some combination of suitably parametrized functions. The main challenge here is to find functions that are (i) easy to parametrize and (ii) can be combined in order to represent the whole waveform. Sigmoids are the most natural choice here, and we are in fact not aware of any other function that would serve our purpose better: Alternative functions like sufficiently high-dimensional Bezier curves or basis functions of splines depend on more than two parameters, which bear no natural relationship to the waveform to be modeled.
% Moreover, a comparison of some alternative approximations to our sigmoidal approximations showed that the latter outperform the former also in terms of accuracy.

Neural networks have been proposed as an alternative for pre-computed tabular gate delay models like CSM and ECSM \cite{ECSM, CCS} already, albeit only for special purposes. In particular, in \cite{Machine_learning}, using a data set constructed by means of SPICE simulations, an ANN was trained to predict the output delay and transition time, based on the input rise time. Its primary purpose was not to replace the way more performant tabular models for simple gates, however, but rather to capture MIS effects in multi-input gates.

\subsection{Paper organization}
\cref{sec:Sigmoidal_Approximation} explains how sigmoids are used to approximate the analog waveforms generated by SPICE simulations. In \cref{sec:Third_Order_Model}, we introduce the particular transfer functions utilized in this paper, which we call \emph{third order model} (TOM). \cref{sec:ANNs} shows how ANNs are used to implement this transfer function. In \cref{sec:Experiments}, some experimental results are presented. \cref{sec:Conclusion} concludes our paper.

\section{Sigmoidal Approximations}
\label{sec:Sigmoidal_Approximation}

According to \cite{mira1995natural}, a sigmoid function is a bounded differentiable real function that is defined for all input values and that has a positive derivative everywhere.

In what follows, we will soley consider sigmoidal approximations based on the variant of the logistics function $\sigma(x)=1/(1+e^{-x})$ given in \cref{eq:single_sigmoid_equation}.

\begin{equation}
\label{eq:single_sigmoid_equation}
  F_s(t,a,b) = \frac{1}{1+e^{-a(t\cdot 10^{10}-b)}}
\end{equation}

In addition to the time parameter $t$ it provides two degrees of freedom via $a$ and $b$: The parameter $a$ controls the \emph{slope}, and consequently the \emph{polarity} (rising/falling), of the transition, while the parameter $b$ gives the absolute time \emph{when} the transition happens. The factor $10^{10}$ accompanying $t$ does not add another degree of freedom, but only helps to keep both parameters in the same range: Since the time periods we work on are in the range of nanoseconds and picoseconds, omitting this multiplicative factor would result in $a$ being in the range of $10^{12}$, while $b$ would be in the range of $10^{-12}$. This would be rather inconvenient for visualization and processing, however.

\subsection{Waveform fitting}

Building on the fact that any bounded real function can be represented as a sum of appropriately parametrized  sigmoids \cite{Cyb89}, we use sigmoids to approximate the signal waveforms observed in digital circuits. In general, such waveforms consist of several transitions. The corresponding sigmoidal approximation allows us to encode a waveform as a list of reals, namely, the parameters of the involved sigmoids. Note that this can be interpreted as some sort of lossy compression.

 In order to determine the parameters
of the sigmoids that approximate a waveform $T$ consisting of $N$ transitions, we use the joint model function $F_T$ defined by
\begin{equation} \label{eq:general_trace_function}
F_{T}(t,a_1,b_1,...,a_{N},b_{N}) = \VDD \sum_{i=1}^{N} F_s(t,a_{i},b_{i})\;,
\end{equation}
where $(a_i, b_i)$ are the parameters of the sigmoid representing the $i$th transition. To obtain the actual parameter values,
the Levenberg-Marquardt least squares fitting algorithm \cite{gavin2019levenberg} is used.  Note carefully that, since adding an arbitrary number of single transition model functions $F_s(t,a_{i},b_{i})$ generally results in a function value between $k\cdot \VDD$ and $(k+1)\cdot \VDD$, the function $F_{T}-k\cdot \VDD$ is actually supplied to the fitting algorithm.

\subsection{Fitting improvements}

Several optimizations to the fitting process, which we applied to SPICE-generated waveforms, had to be implemented to ensure good fitting quality. Since sigmoids cannot model the overshoots/undershoots observed in SPICE waveforms (which are irrelevant for delay estimations anyway, provided they do not extend into the threshold voltage range), see \cref{fig:third_order_model} for example, the waveforms supplied to the fitting procedure are clipped to be in the range of $[0, \VDD]$. Additionally, the weighting vector $\sigma$ offered by the fitting procedure, which offers the possibility to specify a weight to each data point, was used to ensure a tight fit at the inflection points.
% % \todo{maybe talk about initial guesses and bounds?}

\section{Third Order Model Basics}
\label{sec:Third_Order_Model}

We will now introduce the TOM transfer functions considered in this paper. Analogous to digital single-history models for single-input single-output gates like the IDM, recall \cref{sec:intro}, the basic idea is to predict the parameters governing the next output sigmoid of a gate $G$ from the parameters of the current input sigmoid and the previous output sigmoid. Like in the original IDM \cite{ohlinger2021involution}, multi-input gates are reduced to single-input gates by means of internal zero-time boolean gates.

\begin{figure}[t]
  \centering
  \includegraphics[width=1.0\linewidth]{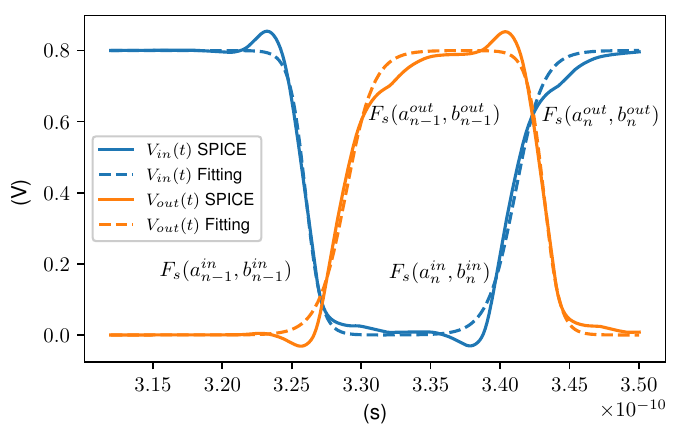}
  \caption{SPICE waveform and fitting of two input transitions and two output transitions of an inverter, along with the parameters expressed in terms of the TOM transfer function.}
  \label{fig:third_order_model}
\end{figure}

Consider the real input resp.\ output waveforms of an inverter depicted by the solid blue resp.\ orange graph in \cref{fig:third_order_model}, each consisting of two transitions, indexed by $n-1$ and $n$. Computing the sigmoidal approximations of these waveforms, as described in \cref{sec:Sigmoidal_Approximation}, leads to an input sigmoid trace represented by $((a_{n-1}^{in}, b_{n-1}^{in})$, $(a_{n}^{in}, b_{n}^{in}))$ and an output waveform represented by $((a_{n-1}^{out}, b_{n-1}^{out})$, $(a_{n}^{out}, b_{n}^{out}))$.

Our actual objective is to find a transfer function $F_G$ that can predict the input-to-output delay
$b_{n}^{out} - b_{n}^{in}$ and the slope of the output transition $a_{n}^{out}$, based on the parameters of the input transition $(a_{n}^{in}, b_{n}^{in})$ and the previous output transition $(a_{n-1}^{out}, b_{n-1}^{out})$:
\begin{equation}
  (a_{n}^{out}, b_{n}^{out} - b_{n}^{in})=F_G(b_{n}^{in}-b_{n-1}^{out}, a_{n}^{in}, a_{n-1}^{out})
  \label{eq:TOM_equation}
\end{equation}
Whereas determining accurate transfer functions is a highly non-trivial problem already for simple gates like inverters, there are
some natural properties that can be exploited here. Among those is time-invariance, i.e., that the prediction does not depend on the absolute timing values $b_{n}^{in}$ and $b_{n}^{out}$ but only their difference. Moreover, the effect of past transitions rapidly decays with time. In particular, if the previous output transition happened sufficiently long ago, it cannot have any significant influence on the prediction of the next output transition, which will hence depend on the parameters of the current input transition only. As already mentioned in \cref{sec:contrib}, our search for reasonably simple but accurate
analytic transfer functions for typical gates like inverters is still on-going.

In this paper, we split the transfer function $F_G$ into $F_{G}^{\uparrow}$ resp.\ $F_{G}^{\downarrow}$ used for predicting the next output transition in the case of a rising resp.\ falling current input transition, and implement those via properly trained ANNs (see \cref{sec:ANNs}). $F_{G}^{\uparrow}$ and $F_{G}^{\downarrow}$ are used by \cref{alg:prediction_base_algorithm}, which computes the prediction of the list of output parameters, given the list of input parameters. It works as follows: Since the list of output parameters is initially empty, it first adds a dummy transition $(s, -\infty)$ to it. Its slope parameter $s$ is set to a fixed value, with the polarity matching the initial conditions of the circuit: If, for example, $G$ is an inverter with initial input tied to $\GND$, $s$ must have a positive sign, as \cref{alg:prediction_base_algorithm} assumes that a rising transition happened at the output at $t=-\infty$. After this initialization, it starts iterating over the input list and computes the list of output transitions, by using either $F_{G}^{\uparrow}$ or $F_{G}^{\downarrow}$ as needed.

\begin{algorithm}[t]
  \KwIn{List of input parameter tuples $((a_{n}^{in}, b_{n}^{in}))_{n \geq 1}$, sorted by ascending $b_{n}^{in}$}
  \KwOut{List of output parameter tuples}
 add $(s, -\infty)$ to Output\;
  Prev $\leftarrow (s, -\infty)$\;
  \While{$(a_{n}^{in}, b_{i_n}^{in}) \in $ \textrm{Input ascending in time} }
  {
	 $(a_{n-1}^{out}, b_{n-1}^{out}) \leftarrow$ Prev\;
	 T $\leftarrow b_{n}^{in} - b_{n-1}^{out}$\;
	 \eIf{$a_{n}^{in} > 0$}
	 {
		$(a_{n}^{out}, \bar{b}_{n}^{out}) \leftarrow F_{G}^{\uparrow}(\textrm{T}, a_{n-1}^{out}, a_{n}^{in})$\;
	 }
	 {
		$(a_{n}^{out}, \bar{b}_{n}^{out}) \leftarrow F_{G}^{\downarrow}(\textrm{T}, a_{n-1}^{out}, a_{n}^{in})$\;
	 }
	 $b_{n}^{out} \leftarrow \bar{b}_{n}^{out} + b_{n}^{in}$\;
	 Prev $\leftarrow (a_{n}^{out}, b_{n}^{out})$\;
	 add $(a_{n}^{out}, b_{n}^{out})$ to Output\;
  }
  \caption{Basic pseudo-code for the output parameter prediction for a single-input gate $G$.}
  \label{alg:prediction_base_algorithm}
\end{algorithm}

The actual prediction algorithm employed in the prototype implementation needs to take care of some aspects that have been ignored in \cref{alg:prediction_base_algorithm}. Most importantly, our basic assumption that each input transition causes exactly one output transition does not always hold. In particular, it is well known that a short input pulse, which barely crosses the threshold voltage, can cause a sub-threshold pulse at the output of a real inverter. This can be taken into account in our algorithm by removing two adjacent tuples that would form such a sub-threshold pulse from the output list. To check whether or not two adjacent tuples $(a_{i-1}^{out}, b_{i-1}^{out})$ and $(a_{i}^{out}, b_{i}^{out})$ form a sub-threshold pulse, one can check whether the sum of the two sigmoids parametrized by $F_S(t,a_{i-1}^{out}, b_{i-1}^{out})$ and $F_S(t,a_{i}^{out}, b_{i}^{out})$ crosses the threshold voltage somewhere. If this is not the case, then the two tuples can safely be dropped from the output list.

% Another complication arises from the handling of multi-input gates, which are reduced to single-input gates that dynamically switch the relevant input.

Gates with more than one input are implemented by one instance of \cref{alg:prediction_base_algorithm} for each input, and a decision procedure that implements the gate logic to determine the currently relevant one. Consider the example of a two input \NOR\ gate: \cref{alg:prediction_base_algorithm} can be performed with input $I_1$ as the relevant one as long as input $I_2=\GND$, i.e., when the last input transition on $I_2$ was a falling one.

\section{Implementing Transfer Functions via ANNs}
\label{sec:ANNs}

In this paper, we use ANNs for implementing the transfer functions $F_{G}^{\uparrow}$ and $F_{G}^{\downarrow}$. Since each of those functions is split into a function that predicts the output slope and another function that predicts the output delay, four ANNs are needed per input.

\begin{figure}[h]
  \centering
  \includegraphics[width=1.0\linewidth]{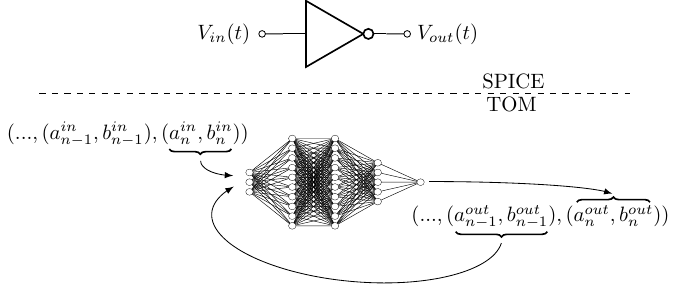}
  \caption{Illustration of the ANN architecture used for each of the 4 transfer functions required for implementing the TOM for an inverter.}
  \label{fig:spice_vs_tom}
\end{figure}

\cref{fig:spice_vs_tom} illustrates the architectural difference between SPICE and our ANN TOM implementation for an inverter: SPICE operates on continuous waveforms and iteratively solves the differential equations that govern currents and voltages numerically. Our TOM implementation uses 4 multilayer perceptron (MLP) ANNs, each consisting of two inner layers with 10 neurons each and a third layer with 5 neurons, with each neuron using a ReLU activation function. The training time of one ANN is less than 10 minutes on a conventional laptop. Note that this architecture was determined by trial-and-error: as it performed very well and could not be substantially improved by slight variations, there was no incentive for a systematic search for an optimal architecture.

\subsection{Training data generation}

In order to train our ANNs (and to generate interpolation polynomials, splines, and look-up-tables for comparison purposes), we needed a sufficiently large set of representative input-to-output relations,
for every elementary gate $G$ (inverter, \NOR) supported by our simulator. They have been obtained
by systematic SPICE simulation using the $\SI{15}{\nano\meter}$ FinFet models of the Nangate Open Cell Library \cite{Nangate15nmLibrary} at $V_{DD} = \SI{0.8}{\volt}$, which we explain below by the example of a \NOR\ gate.

\begin{figure}[h]
  \centering
  \includegraphics[width=1.0\linewidth]{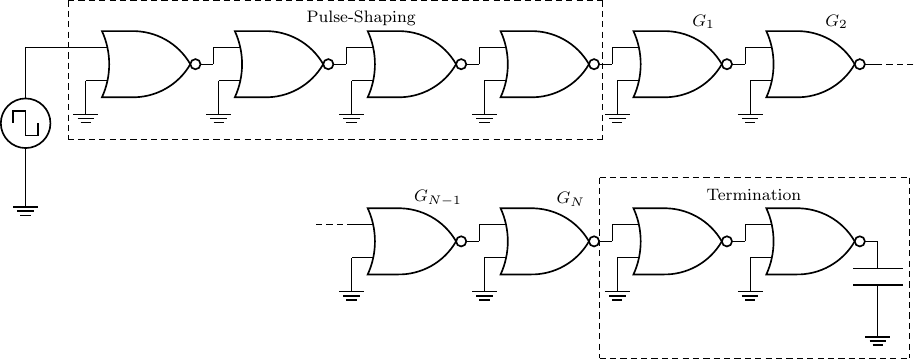}
  \caption{Example \NOR\ chain, stimulated by systematically varied input pulses, for generating training data.}
  \label{fig:data_generation_circuit}
\end{figure}

Consider \cref{fig:data_generation_circuit}, which shows a chain of \NOR\ gates, stimulated by a voltage source capable of generating traces of Heaviside transitions in a carefully controlled way. The first few \NOR\ gates are used for pulse shaping purposes, which convert the artificial Heaviside transitions into realistic input waveforms that are supplied to the identical target gates $G_1$ to $G_N$ to be modeled. Similarly, a few \NOR\ gates appended at the end ensure a proper termination of the outputs of the target gates. Together, these parts of the circuit ensure that the waveforms obtained from $G_1$ to $G_N$ are not biased by their position in the chain. Interconnecting parasitics are identical for all stages and not shown in the schematics.

\begin{figure}[h]
  \centering
  \includegraphics[width=1.0\linewidth]{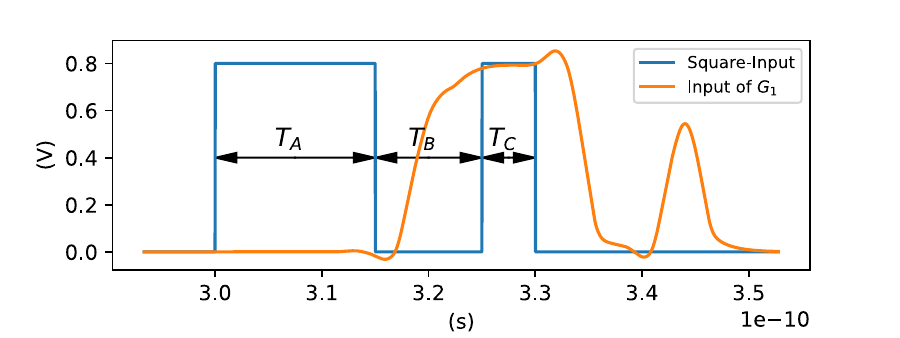}
  \caption{Heaviside input pulses and pulse-shaped input of the first gate $G_1$.}
  \label{fig:square_input_stimulus}
\end{figure}

The input of the chain is stimulated by four Heaviside transitions, governed by the three time intervals $T_A, T_B$ and $T_C$, as shown in \cref{fig:square_input_stimulus}.

As these transitions traverse the entire chain, the waveforms of $G_1$ to $G_N$ are recorded and taken as $N$ independent data for training a single \NOR\ gate. By employing the fitting algorithm described in \cref{sec:Sigmoidal_Approximation} on the waveforms at the input and output of $G_1$ to $G_N$, we
obtain a table of input-to-output parameter relations needed for (training) the transfer function
\cref{eq:TOM_equation}. To cover all reasonably possible fitting parameter combinations, we systematically
varied $T_A, T_B$ and $T_C$ within the range between $\SI{5}{\pico\second}$ and $\SI{20}{\pico\second}$ with a granularity of $\SI{1}{\pico\second}$, which corresponds to approximately $15^3$ different SPICE simulation
runs. \Cref{fig:data_generation_circuit} is only one of several circuits that were used to generate training data. For example, circuits with \NOR\ gates having a fan-out of two were also part of this set of circuits.

 \subsection{Valid region containment}

One inherent problem of any approach based on approximations, including sigmoidal approximations, is error amplification: Since the output prediction of some gate $G$ serves as the input of the successor gate $G'$ in a circuit, prediction errors caused by the transfer function of $G$ will usually increase the error caused by the transfer function of $G'$. Since ANNs can show arbitrary behavior if they are evaluated on inputs that are far outside the training set, we added a mechanism that circumvents this problem.

Since the set of input parameters ($b_{n}^{in}-b_{n-1}^{out}$, $a_n^{in}$, $a_{n-1}^{out}$) of a training set populate three dimensional Euclidean space only in certain region, input parameters that lie outside of this valid region can be detected. By computing the concave hull of this set, we can generate a region that distinguishes valid input parameters from input parameters that accumulated too much error. If the input supplied to an ANN (after training) lies outside of this concave hull, we compute the closest point on the concave hull and use these coordinates as inputs instead. If the input parameters already lie inside the concave hull nothing needs to be done. We stress that computing the concave hull of a set is a delicate task, however, since unlike the convex hull, it is generally not uniquely defined \cite{concavehull}.

% \begin{figure}[h]
%   \centering
%   \includegraphics[width=0.75\linewidth]{graphics/matlab_parameters_cropped}
%   \caption{Input parameters populating three dimensional space for a \NOR\ gate.}
%   \label{fig:input_parameters_in_3d_space}
% \end{figure}

\section{Experiments}
\label{sec:Experiments}
In order to demonstrate the principal feasibility of our approach, we implemented a simple and very restricted prototype of a simulator, which can be used for simulating circuits made up of inverters and \NOR\ gates. We simulated the instances c17, c499 and c1355 of the ISCAS-85 Benchmark \cite{ISCAS85_reference} in SPICE (actually, Spectre version 20.1), in ModelSim (version 10.5), and in our prototype simulator. Whereas the results of our experiments are by no means meant to be representative, they nevertheless demonstrate the potential of our approach.

\subsection{Prototype simulator}
Our prototype simulator, which is available on github, \cite{prototype_github} allows to simulate arbitrary circuits consisting of elementary gates. Input signals are supplied to the simulator in the form of sigmoid parameter lists. In its current version, only \NOR\ and inverters are supported as elementary gates, and all elementary gates of the same type are identical, i.e., are modeled by the same transfer function (ANN). The only exception are \NOR\ gates with fan-out of 2 or more, which use different ANNs than \NOR\ gates with fan-out 1.

\subsection{Experimental setup}
Since the ISCAS-85 circuits consist of gates of any type, whereas our prototype simulator only supports inverters and \NOR\ gates so far, we simulate ISCAS-85 circuits where each non-\NOR\ gate is replaced by an equivalent circuit consisting of just \NOR\ gates. Since \NOR\ gates are functionally complete, this is always possible. As this replacement inflates the number of gates a circuit is made of, \cref{tab:benchmark_table} lists the number of \NOR\ gates for each circuit.

For the ModelSim simulations, the parasitics and delay values were extracted using the Cadence tools Genus and Innovus (version 19.11). As we do not currently provide trained ANNs for \NOR\ gates with different parasitics and fan-outs larger than 2, all our SPICE simulations were conducted for circuits that have the same interconnect as used for our gate characterization.

The inputs of all circuits were stimulated by randomized transition sequences with inter-transition times having a normal distribution, given by $\mu_t,\sigma_t$. All circuits were stimulated with three different random sets of stimuli $(\mu_t,\sigma_t)=(20\si{\pico\second},10\si{\pico\second})$, $(100\si{\pico\second},50\si{\pico\second})$ and $(500\si{\pico\second},250\si{\pico\second})$, with the number of transitions being 20, 10 and 5. For each setup 50 randomized runs were performed. As in the case of the training circuits, e.g., \cref{fig:data_generation_circuit}, the SPICE circuits were augmented by pulse-shaping at the inputs and termination at the outputs, which, of course, slightly increases simulation time.

In order to numerically compare the overall prediction performance of ModelSim and our approach (see \cref{tab:benchmark_table}), the outputs of both ModelSim and our simulator were compared to the respective SPICE simulation. The total amount of time $t^{err}_{ModelSim}$ and $t^{err}_{Sigmoid}$ during which the respective prediction and SPICE did not match were summed among all outputs of a circuit and compared. In this comparison, the prediction trace and the SPICE trace are considered to match at time $t$ if both traces are above(below) the threshold $\frac{V_{dd}}{2}$.

\subsection{Experimental results}

\begin{table*}[t]
\caption{Numerical results of our experiments.}
\begin{center}
\begin{tabular}{|c|c|c|c|c|c|c|c|}
\hline
ISCAS85 circuit&\#\NOR-gates&$\mu_t,\sigma_t (\si{\pico\second})$&error ratio&$t^{err}_{ModelSim}(\si{\pico\second})$&$t^{err}_{Sigmoid}(\si{\pico\second})$&$t^{sim}_{Sigmoid}(\si{\second})$&$t^{sim}_{SPICE}(\si{\second})$\\
\hline
\multirow{3}{*}{c17}&\multirow{3}{*}{24}&\phantom{0}20,10\phantom{0}&0.44&45.79&20.00&\phantom{0}6.0&\phantom{00}22(\phantom{0}108)\\
\cline{3-8}
&&100,50\phantom{0}&0.76&23.99&18.22&\phantom{0}6.3&\phantom{00}19(\phantom{00}93)\\
\cline{3-8}
&&500,250&0.88&11.99&10.53&\phantom{0}7.2&\phantom{00}13(\phantom{00}64)\\
\hline
\multirow{3}{*}{c499}&\multirow{3}{*}{860}&\phantom{0}20,10\phantom{0}&0.36&23.80&\phantom{0}8.65&13.3&\phantom{0}739(4603)\\
\cline{3-8}
&&100,50\phantom{0}&0.83&\phantom{0}9.94&\phantom{0}8.28&17.4&\phantom{0}570(3696)\\
\cline{3-8}
&&500,250&1.29&16.99&22.02&12.9&\phantom{0}800(4879)\\
\hline
\multirow{3}{*}{c1355}&\multirow{3}{*}{2068}&\phantom{0}20,10\phantom{0}&0.14&75.28&10.21&19.8&1210(8159)\\
\cline{3-8}
&&100,50\phantom{0}&0.34&23.23&\phantom{0}7.87&25.1&1037(6900)\\
\cline{3-8}
&&500,250&0.94&24.73&23.12&19.3&\phantom{1}969(6703)\\
\hline
c1355(same stimulus)&2068&\phantom{0}20,10\phantom{0}&0.50&75.28&37.46&19.5&1210(8159)\\
\hline
\end{tabular}
\end{center}
\label{tab:benchmark_table}
\end{table*}

\Cref{tab:benchmark_table} summarizes the main numerical results of experiments: $t^{err}_{ModelSim}$ and $t^{err}_{Sigmoid}$ represent the average amount of time the prediction does not match SPICE, the ratio $t^{err}_{Sigmoid}/t^{err}_{ModelSim}$ is given in a dedicated column. $t^{sim}_{Sigmoid}$ and $t^{sim}_{SPICE}$ provide the average simulation time, i.e., wall clock time. Since Spectre supports multi-threading and used up to 8 threads per SPICE simulation, the total CPU time is appended in parentheses. ModelSim simulation time is not listed explicitly, as it was below $1\si{\second}$ in all cases.

Inspecting \Cref{tab:benchmark_table}, it is apparent that our simulator provides significantly less error than ModelSim. Especially for short inter-transition times, it has up to 86\% less error than ModelSim in the case of c1355 and $\mu_t = 20\si{\pico\second} ,\sigma_t=10\si{\pico\second}$. For all three circuits, the advantage of our simulator becomes smaller as inter-transition times increase, however. This is particularly pronounced in the case of $\mu_t = 500\si{\pico\second},\sigma_t=250\si{\pico\second}$, where small inter-transition times (which are handled well by our simulator, but poorly by ModelSim) are very rare. This apparent anomaly is a consequence of the much more accurate gate characterization used for ModelSim, which cannot be matched by our just two fan-out 1 and fan-out 2 \NOR\ gate ANNs. After all, the ModelSim simulations are based on individual gate delay models for each gate in a circuit, which also include the particular interconnect delay. On the other hand, our simulator currently applies the same FO1 ANNs to each \NOR\ gate with fan-out 1 and the same FO2 ANNs to any \NOR\ gate with a fan-out of 2 or more.

Since the engineering level and maturity of our prototype simulator is very low, compared to Spectre and ModelSim, it is not very meaningful to directly
compare the simulation times. It is revealing, though, that our simulator nevertheless outperformed Spectre by up to a factor of 60 for the circuit c1355 in terms of wall time, which corresponds to a factor of around 400 in terms of total CPU time.

\begin{figure}[h]
  \centering
  \includegraphics[width=1.0\linewidth]{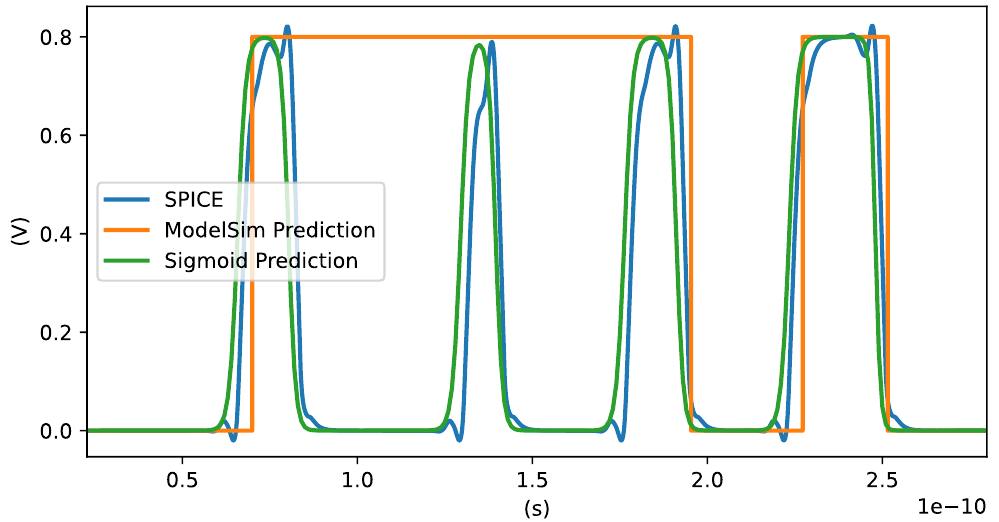}
  \caption{Example ModelSim and sigmoid prediction compared to the respective SPICE simulation.}
  \label{fig:example_benchmark_trace}
\end{figure}

In addition, a detailed trace-based comparison between ModelSim and our tool was made in the case of c1355 with $\mu_t=20 \si{\pico\second},\sigma_t=10 \si{\pico\second}$ (last column of \cref{tab:benchmark_table}). For the purpose of this comparison, our sigmoid simulator was stimulated with exactly the same input waveforms as ModelSim. \cref{fig:example_benchmark_trace} depicts an example output trace showing the prediction of ModelSim, our sigmoid prediction, and the SPICE reference. It clearly shows the accuracy advantage of our simulator compared to ModelSim. For example, ModelSim predicts only one rising and one falling transition between $50\si{\pico\second}$ and $200\si{\pico\second}$, whereas our simulator predicts three rising and three falling transitions, in accordance with SPICE. Even though the transitions predicted by our simulator are too early on average (which is again a consequence of our inferior gate characterization compared to ModelSim), the absolute error time is significantly less than ModelSim: on average, our simulator provided 50\% less error than ModelSim in this direct comparison.

\section{Conclusions}
\label{sec:Conclusion}

We advocated a generalization of dynamic timing analysis of digital circuits based on signal traces consisting of a sequence of digital (Heaviside) transitions to traces consisting of a sum of time-shifted sigmoids. Gates are described by means of transfer functions here, which describe how the sigmoid parameters, namely, occurrence time and slope, of an output trace depend on the sigmoid parameters of the relevant input traces. We developed a preliminary prototype simulator (available at github) \cite{prototype_github} for circuits made up of \NOR\ gates and inverters, which implements their transfer functions via ANNs. Since it runs substantially faster than an analog simulator, while offering better accuracy than a digital simulator, it showcases
the potential of our approach.

Part of our current/future work is devoted to extend our simulator by ANNs for elementary gates with arbitrary fan-in, fan-out, loads and parasitics, which would make it more interesting from a practical perspective. Moreover, our simulator is also used for guiding our attempts to develop \emph{analytic} transfer functions for elementary gates like inverters and \NOR.

\bibliographystyle{IEEEtran}
\bibliography{citations}

\end{document}